\pdfoutput=1
\documentclass[preprint,preprintnumbers,amsmath,amssymb,superscriptaddress]{revtex4}
 
\newcommand{\be}{\begin{equation}}
\newcommand{\ee}{\end{equation}}
\newcommand{\ba}{\begin{eqnarray}}
\newcommand{\ea}{\end{eqnarray}}
\newcommand{\bd}{\begin{displaymath}}
\newcommand{\ed}{\end{displaymath}}

\newcommand{\commentout}[1]{{}}

\usepackage{float}
\usepackage{amssymb}
\usepackage{color}
\usepackage{slashed}
\usepackage{amsmath}
\usepackage{graphicx}
\usepackage{epstopdf}
\usepackage{dcolumn}
\usepackage{bm}
\usepackage{hyperref}

\makeatletter
\renewcommand\thefigure{\@arabic\c@figure}
\renewcommand\fnum@figure{\figurename~\thefigure}
\newcommand\onlinecap{\renewcommand\fnum@figure{\figurename~\thefigure~(Color Online)}}
\makeatother

\begin{document}

\title{Dissipative effects on quarkonium spectral functions}
\author{Yusuf Buyukdag}
\email{buyuk007@umn.edu}
\affiliation{School of Physics \& Astronomy, University of Minnesota, Minneapolis, MN 55455,USA}
\author{Clint Young}
\email{young@physics.umn.edu}
\affiliation{School of Physics \& Astronomy, University of Minnesota, Minneapolis, MN 55455,USA}
\date{\today}

\begin{abstract}

Quarkonium at finite temperature is described as an open quantum system whose dynamics are determined by a potential $V_R({\bf x})$ and drag coefficient $\eta$, using a path integral with a non-local term. Path-integral Monte Carlo calculations determine the Euclidean Green function for this system to an accuracy greater than one part in a thousand and the maximum entropy method is used to determine the spectral function; challenges facing any kind of deconvolution are discussed in detail with the aim of developing intuition for when deconvolution is possible. Significant changes to the quarkonium spectral function in the $1S$ channel are found, suggesting that any description of quarkonium at finite temperature, using a potential, must also carefully 
consider the effect of dissipation.

\end{abstract}

\maketitle

\section{Introduction}

At both zero temperature and above deconfinement, quarkonium is an ideal probe of QCD. At zero temperature its mass $M \gg \Lambda_{ {\rm QCD} }$ suggests the effective field theory approach of NRQCD [1], where the heavy quark mass is integrated out. In effect, the heavy quark mass sets the momentum scale of QCD for states using this description, allowing perturbative results to work well. The situation becomes more complicated at finite temperatures. NRQCD can be extended to finite temperature and used to examine the break-up of quarkonium \cite{Brambilla:2013dpa}.

An alternative point of entry is to treat quarkonium above deconfinement as an open quantum system (OQS)\cite{Young:2010jq}. This departs from any attempt to describe quarkonium with 
perturbation theory, and the parameters describing quarkonium have to be determined separately, either in calculations based on first principles or from experimental measurements. However, 
this description has the strength of being independent of some of the scale hierarchies. One would suspect that the OQS approach is at its most useful for charmonium and highly excited states of 
bottomonium, where the binding energy is small and the confining term of the Cornell potential is important for describing the spectrum of states at zero temperature. Perhaps the 
greatest strength of the OQS approach is that the approach to thermalized yields of quarkonium is completely natural, which is not necessarily the case for some potential models.
Since the first, simplest descriptions of quarkonium in this way, the treatment of heavy quarks at finite temperature has developed significantly, thanks the work of  several authors 
\cite{Beraudo:2010tw, Akamatsu:2014qsa, Akamatsu:2011se}.

This paper describes the numerical determination of the spectral function for an open quantum system, which has the potential term and the diffusion coefficient matched to quarkonium at the 
temperatures reached in heavy-ion collisions. The procedures used can be generalized for different densities of states for the thermal bath, for different couplings, and for non-trivial 
correlations of the bath's force on the heavy quark and the anti-quark. In Section \ref{correlators}, the path-integral Monte Carlo algorithm for this open quantum system is described. In Section 
\ref{deconvolution}, intuition is developed for when phenomenologically significant deconvolution of spectral functions is possible, with this having some implications for results using lattice QCD calculations; and finally, the maximum entropy method is used to deconvolve the results from Section \ref{correlators} into quarkonium spectral functions, and the non-trivial relationship between 
diffusion and the destruction of the $J/\psi$ state can be determined. In this way, the often competing effects of drag and momentum diffusion on $J/\psi$ survival rates can be simultaneously 
considered, as was done in \cite{Young:2008he, Young:2011ug}.

While this procedure may seem cumbersome, it is in fact relatively cheap computationally. The most costly step is the calculation of the correlation functions to sufficient accuracy so that deconvolution yields significant results; work on high-order estimates of the action in Equation \ref{Eq:Gtau} can speed this computation significantly as it has for simple actions.

\section{Euclidean current-current correlators and dissipative effects}
\label{correlators}

Much of the groundwork towards examining open quantum systems with path integrals was already done by Feynman and Hibbs when they considered influence functionals 
\cite{Feynman:1965qm}; Caldeira and Leggett used path integrals to describe a system with a specific thermal bath \cite{Caldeira:1982iu} while Grabert, Schramm, and Ingold generalized Caldeira and Leggett's 
results \cite{Grabert:1988yt}. In \cite{Young:2010jq}, these techniques were applied to imaginary-time Green functions. To review: one starts with the action for a heavy degree of freedom with position $x$ interacting with a light degree of freedom described by $R$. This light degree of freedom is often a collective mode of a gas or of condensed matter, and not just a value for position. The action for this system, using a harmonic approximation and minimal coupling, is 
\begin{equation}
S=\int_0^{\tau} d\tau^{'} \left[\frac{1}{2}M\dot{x}^2+V(x)
+\frac{1}{2}m\dot{R}^2+\frac{1}{2}m\omega^2 R^2 -CxR \right] \nonumber
\end{equation}
gives the influence functional 
\begin{align}
\left \langle x_f, \tau | x_i, 0 \right \rangle_{{\rm red}}
& = 
\int {\cal D}x\; \exp \left( -\int_0^{\tau}d\tau^{'} \left[\frac{1}{2}M\dot{x}^2+V(x) \right. \right. \nonumber \\
 & \left. \left. -\frac{C^2}{2m\omega \sinh(\omega \tau)}x(\tau^{'}) \cosh(\omega(\tau-\tau^{'}))
\int_0^{\tau^{'}}ds\; x(s)\cosh(\omega s)\right] \right) {\rm .}
\label{IF} \\ \nonumber
\end{align}
after finding $\left\langle x_f, R_f | x_i, R_i \right\rangle$ and integrating over $R_i$ and $R_f$. When there are multiple light degrees of freedom, the final term in the integral becomes the  
sum $\sum_i \frac{C^2_i}{2m_i\omega_i \sinh(\omega_i \tau)}x(\tau^{'}) \cosh(\omega_i(\tau-\tau^{'}))
\int_0^{\tau^{'}}ds\; x(s)\cosh(\omega_i s)$. The system becomes dissipative in the limit of an infinite number of light degrees of 
freedom of masses $m_i$ and frequencies $\omega_i$: the density of states 
\begin{equation}
C^2(\omega)\rho_D(\omega) = \begin{cases} 
\frac{2m\eta\omega^2}{\pi}  & \text{if } \omega < \Omega\\ 
0 & \text{if } \omega > \Omega \end{cases}
\label{densityOfStates}
\end{equation}
in the limit $\Omega\to\infty$ was examined in \cite{Caldeira:1982iu} and found to lead to a path integral which yields the Langevin equation in the classical, high-temperature limit. Performing the integral over these states in Equation \ref{IF} leads to the integral  
\begin{align*}
\setlength{\jot}{12pt}
& \int_0^{\Omega}d\omega \int_0^{\tau} du \int_0^{\tau}dv\; x(u) x(v) \frac{\omega \cosh(\omega (\tau-u)) \cosh(\omega v)}{\sinh(\omega \tau)}
\theta (u-v) \\
&= \frac{\Omega}{2}\int_0^{\tau} du(x(u))^2-\int_0^{\Omega}\frac{d\omega}{\omega \sinh(\omega \tau)} 
\int_0^{\tau} du\int_0^u dv\; \dot{x}(u) \dot{x}(v) \sinh(\omega(\tau-u)) \sinh(\omega v) {\rm ,}\\
\end{align*}
which forces a renormalization of $V(x)$, and gives the imaginary-time Green function for this system:
\begin{equation}
\label{Gtau}
\hspace{-.5cm}
G_{ {\rm red} }(x_f, x_i, \tau) = \int {\cal D}x\; \exp \left( -\int_0^{\tau}du \left[\frac{1}{2}M\dot{x}(u)^2+V_R(x(u)) -\frac{\eta}{2\pi} \int_0^u dv \; \dot{x}(u) \dot{x}(v) \log \left[
  \frac{\sin(\frac{\pi}{2} \frac{u-v}{\tau})}{\sin(\frac{\pi}{2} \frac{u+v}{\tau})} \right] \right] \right){\rm .} \nonumber
\end{equation}
Note that only the potential is renormalized, and that the final term introduced by the integration over the bath of particles is translationally invariant, meaning that no 
finite, $x$-dependent term has been added to the path integral. This imaginary-time Green function 
can be made periodic in $\tau$ with period $\beta$ using the method of images to make a quantity related to results from finite-temperature lattice calculations.

For a general potential, numerical methods for determining this path integral must be developed. Path-integral Monte Carlo techniques have been developed for condensed matter systems; 
Ceperley reviewed these methods as they concern liquid helium \cite{Ceperley:1995zz}. For quarkonium, a relatively simple algorithm is used: the path is discretized to $2^{11}+1$ equally-spaced times, and free-particle paths are sampled using the bisection method. These paths are reweighted according to the expression above. The double integral in the exponential poses some problems because it is improper; both the ``square" and ``triangular" regions in the double integral must have their measures determined analytically so that sufficient precision may be achieved.

\begin{figure}
 \includegraphics[width=5in]{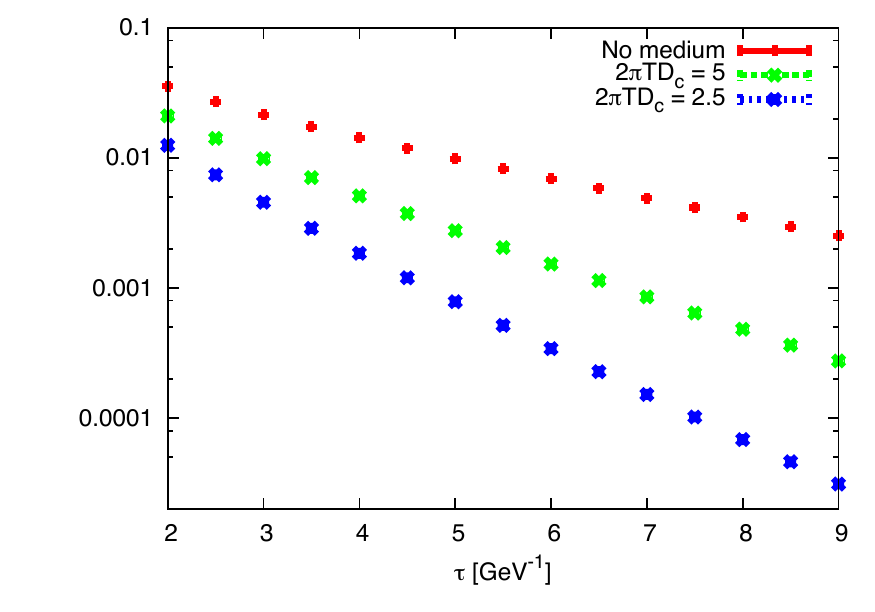}
\onlinecap
\caption{From \cite{Young:2010jq}, $G(\tau)$ with and without dissipative effects.
}
\label{G_tau_tauShift2}
 \end{figure}

The results for $G(\tau)$ have been calculated for the Cornell-like potential
\begin{equation}
V(r) = \begin{cases}
	-\frac{1.5\alpha_s}{r_{{\rm min}} }+ 0.5\sigma r_{{\rm min}} + (\frac{0.5\alpha_s}{r_{{\rm min}} }+ 0.5\sigma r_{{\rm min}})\frac{r^2}{r_{{\rm min}}^2} \\
	 \text{ if }r<r_{{\rm min}}{\rm ,} \\
	-\alpha_s/r+\sigma r  \text{ if }r>r_{{\rm min}} \\
	\end{cases}
\end{equation}
with $r_{{\rm min}}$ chosen to be small, $0.4\; {\rm GeV}^{-1}$ (here, all units are in GeV), $\alpha_s = 0.499$, and $\sigma = 0.16\;{\rm GeV}^2$. The purpose of using this piecewise function 
is to simplify dealing with the divergence of the Cornell potential as $r \to 0$. The drag coefficients $\eta=$0, 0.0729 GeV, and 0.1458 GeV were used, corresponding to the spatial diffusion coefficients 
$2\pi D_c=\infty$, 5, and 2.5, respectively, at $T=285\;{\rm MeV}$. The results for these values were shown in \cite{Young:2010jq}; here, in Figure \ref{G_tau_tauShift2}, we show results for the same values but for a large range in $\tau$, which is necessary for extracting the spectral function. We should also note that some degrees of freedom of the heavy quark are not represented in this path integral,
namely, spin and color. The potential term is therefore the thermal average of this quantity over spin and color states. In \cite{Akamatsu:2014qsa}, Lindblad equations describing the evolution of color 
singlet and octet states are determined for heavy quarkonium, starting from perturbative QCD for heavy quarks. Our path integral, on the other hand, requires color-averaged potentials as an input for 
the path integral Monte Carlo calculations.

\section{The quarkonium spectral function and the challenges facing deconvolution}
\label{deconvolution}

The imaginary-time Green function is related to the spectral function $\rho = -{\rm Im} \{ G_R \}$ through a Laplace transform:
\begin{equation}
\label{Eq:Gtau}
G(\tau) = \int \exp(-\omega \tau) \rho(\omega) d\omega{\rm ;}
\end{equation}
similarly, the imaginary-time finite-temperature Green function is this Laplace transform made periodic through the method of images:
\begin{equation}
\label{Eq:Gtaubeta}
G(\tau, \beta) = \int \frac{ \cosh(\omega(\tau-\beta/2))}{ \sinh(\omega \beta/2) } \rho(\omega) d\omega{\rm .}
\end{equation}
Complex analysis helps here: the Fourier transforms of the various Green functions are conveniently related to each other; as a function of complex 
$\omega$, causality translates to analyticity of $G_R$ in the upper-half plane; and the relation between $G_R$ and $G_A$ in $\omega$ is a generalization of the fluctuation-dissipation 
theorem \cite{Son:2009vu}.

\begin{figure}
 \includegraphics[width=5in]{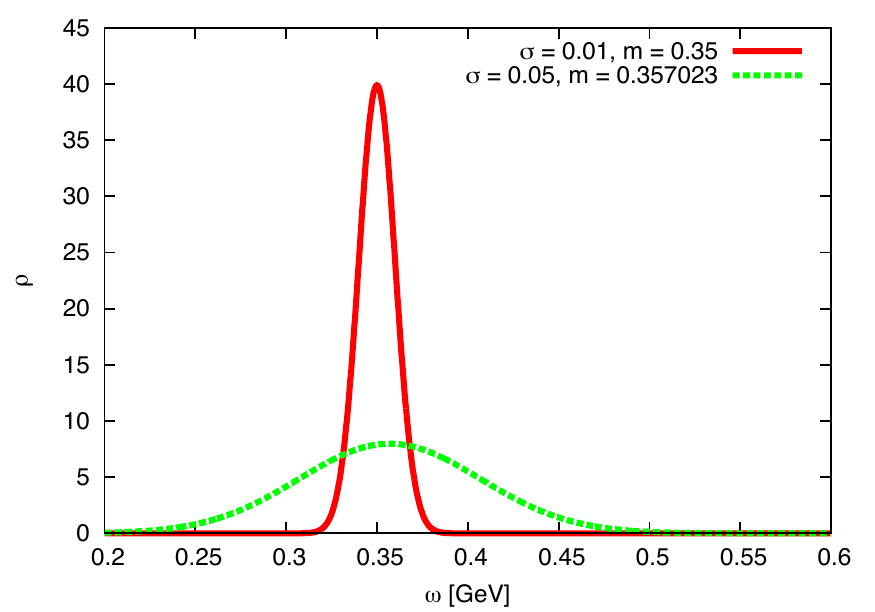}
\onlinecap
\caption{The spectral functions $\rho_1$ (with width $\sigma=0.01\;{\rm GeV}$) and $\rho_2$ (with width $\sigma=0.05\;{\rm GeV}$).
}
\label{twoRhos}
 \end{figure}

Unfortunately, when working with numerical results and not analytic expressions for $G(\tau)$, these results are not helpful. The reason for this is simple: significantly different spectral functions often differ by very little in their Laplace transforms. This is not obvious to theorists who work with analytic results so it is necessary to illustrate this with an example: consider two spectral functions, $\rho_1$ and $\rho_2$, plotted in Figure \ref{twoRhos}. The widths differ by a factor of five, from $0.01\;{\rm GeV}$ for $\rho_1$ to $0.05\;{\rm GeV}$ for $\rho_2$. This changes the lifetime of the state from $\sim 20\;{\rm fm/c}$ to $\sim 4\;{\rm fm/c}$. The lifetime of the state represented by $\rho_1$ is long compared with the timescales of the heavy-ion collisions at RHIC and the LHC; it represents a state whose yields would be largely unaffected, while the state represented by $\rho_2$ would be significantly suppressed.

\begin{figure}
 \includegraphics[width=5in]{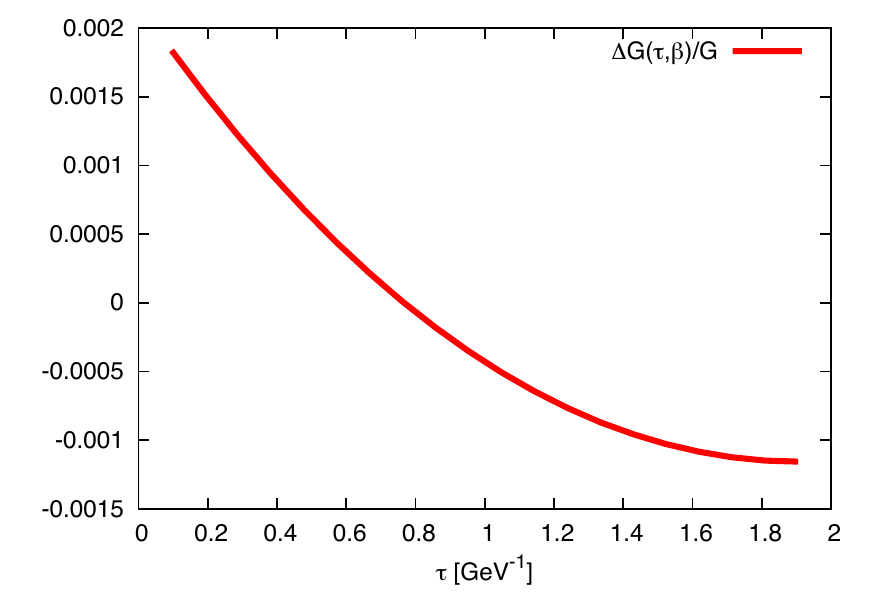}
\onlinecap
\caption{The relative difference $\Delta G/G=(G_2(\tau,\beta)-G_1(\tau,\beta))/G_2(\tau,\beta)$ between the Green functions obtained from $\rho_1$ and $\rho_2$.
}
\label{DG}
 \end{figure}

Lattice QCD calculations of quarkonium correlation functions have determined $G(\tau, \beta)$ for temperatures between $T_c=175\;{\rm MeV}$ and $2T_c$, by determining the correlation functions of composite operators
\begin{equation}
G(\tau, \beta) = \int d^3x \left \langle J_{\mu}({\bf x}, \tau) J^{\mu}({\bf x}, \tau) \right \rangle {\rm ,}
\end{equation}
where $J_{\mu} = \bar{\psi}\Gamma_{\mu} \psi$ and $\Gamma_{\mu} = \gamma_{\mu}{\rm ,}\; \gamma_4 \gamma_{\mu}$. This function corresponds to the transform of the spectral function shown in Equation \ref{Eq:Gtaubeta}. For $\beta = (1.5T_c)^{-1}$, this transform of $\rho_1$ and $\rho_2$ can be computed; Figure \ref{DG} shows the relative difference $\Delta G/G \equiv 
(G_2-G_1)/G_2$ for values of $\tau$ from zero to $0.5\beta$.

The relative difference has a maximum of about one part in a thousand. If the full range for $\tau$ were equally sensitive to changes in this bound state, this would suggest that  three results for 
$G(\tau, \beta)$, equally spaced on this range, would need an accuracy of a few parts in ten thousand to determine which spectral function fits the results best. Constraining the spectral function in the vector channel with sum rules is not likely to help much because 
they provide only a few constraints to a continuous function.

\begin{figure}
 \includegraphics[width=5in]{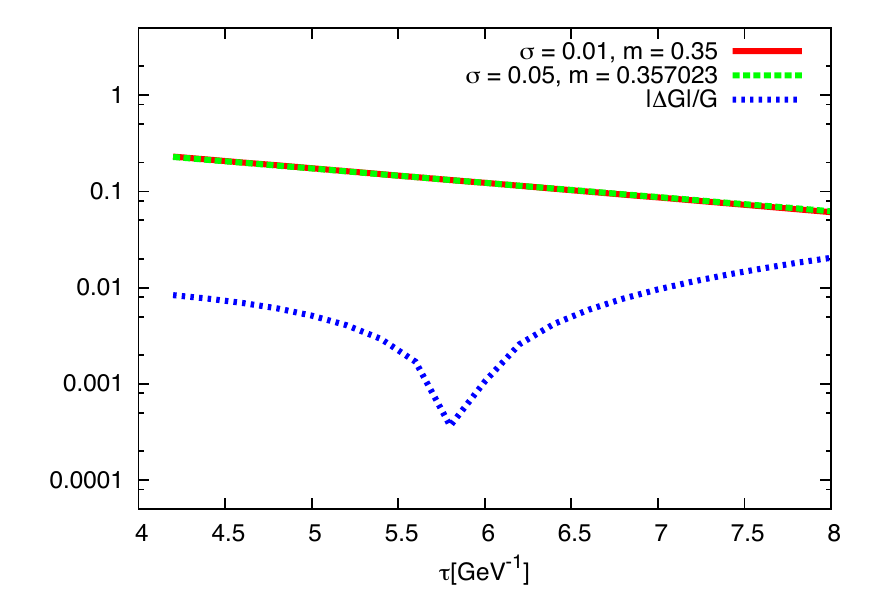}
\onlinecap
\caption{$G_1(\tau)$, $G_2(\tau)$, and $|\Delta G|/G$, plotted together.
}
\label{G}
 \end{figure}

Lattice QCD calculations at specific temperatures are performed on periodic lattices and are used to determine finite-temperature Green functions as in Equation \ref{Eq:Gtaubeta}, not 
\ref{Eq:Gtau}. However, if another calculation can determine imaginary-time Green functions as in Equation \ref{Eq:Gtau} at large $\tau$, it will be sensitive mostly on the shape of the ground state 
and the resulting $G_1$ and $G_2$ should differ by a greater relative $\Delta G$. We take advantage of this fact by calculating this Green function for a dissipative system over the range shown 
in Figure \ref{G}. Here, $|\Delta G|/G$ has a maximum of about 1\%, suggesting that numerical deconvolution will be less daunting.

If $G(\tau)$ is determined with sufficient accuracy, deconvolution can be attempted. Again, the inverse Laplace transform is analytically easy but numerically non-trivial. Working with tens of data 
points for $G(\tau)$, no matter the accuracy, underdetermine any reasonable discretization of the spectral function. Consider $\chi^2$-minimization in this situation: the spectral function is discretized into $\sim 1000$ points and $\sim 10$ data points are fitted, unnaturally small values of $\chi^2$ can be achieved. The result for $\rho(\omega)$ usually appears choppy in this 
situation. The principle of maximum entropy suggests that the best fit for $\rho(\omega)$ is not the fit with the smallest $\chi^2$, but the fit with a reasonable value for $\chi^2$ but is also constrained by 
the information entropy
\begin{equation}
I = \sum_i \left[ \rho_i\log\left(\rho_i/\sigma_i \right) - (\rho_i-\sigma_i) \right]{\rm ,}
\end{equation}
where $\rho_i$ are the discretized values for $\rho(\omega)$ and $\sigma_i$ are the {\it priors} for $\rho_i$. This constraint on the $\chi^2$-minimization is often cast in terms of Bayesian 
principles, but it can also be thought of as an assumption of smoothness for the deconvolved function, far less controversial in physical situations. The sum of $\chi^2$ and $I$ define the 
energy function to be minimized:
\begin{equation}
E(\rho_i) = \chi^2(\rho_i)+\alpha I {\rm .}
\end{equation}
The coefficient $\alpha$ determines the relative importance of $\chi^2$ and $I$ in the minimization; it is often chosen to make $\chi^2$ roughly equal the number of data points. Gallicchio and 
Berne point out that the best value for $\alpha$ can also be determined with Bayesian logic \cite{Gallicchio:1996me}. Our choices for $\alpha$ will be based simply on whether or not they yield reasonable 
values for $\chi^2$.

A suitable algorithm for the minimization of this function is simulated annealing: 1.) random steps are considered in the space of values for $\rho_i$, , changing the value of $E$ from 
$E_i$ to $E_f$, 2.) any step that decreases $E$ is taken while any step increasing $E$ is taken with probability $P=\exp(-(E_f-E_i)/T)$, $T$ being a temperature chosen to be large initially, and 
3.) the process is repeated with $T$ lowered until a minimum temperature $T$ is reached. When a function is multi-modal as is possibly the case, simulated annealing is far more likely 
to be successful at finding absolute minima than the biconjugate gradient method, or any other slope-following method.

At this point, a test of this method of deconvolution would be useful. We can test this by going in the opposite direction: starting with a given spectral function, finding its Laplace 
transform (with random Gaussian error added), and deconvolving with the maximum entropy method. The spectral function  
\begin{equation}
\begin{split}
\sigma (\omega) & = 1.2 (e^{-\frac{(\omega - 0.33)^2}{2*0.0174^2}} - e^{-\frac{(\omega + 0.33)^2}{2*0.0174^2}} ) \\ & + 0.2 (e^{-\frac{(\omega - 1)^2}{2*0.108^2}} - e^{-\frac{(\omega + 1)^2}{2*0.108^2}} )
\end{split}
\end{equation}
mimics the form that we expect for the quarkonium spectral function at finite temperature. The data set 
\begin{equation}
G(\tau_i) = \int \sigma (\omega) e^{-\omega \tau} d\omega + \Delta G_i
\end{equation} 
contains 15 points ranging from $\tau = 0.5\;{\rm GeV}^{-1}$ and $\tau = 9.6\;{\rm GeV}^{-1}$, and $\Delta G_i$ is random Gaussian noise with standard deviation $10^{-6}\;{\rm GeV}$.
In Figure \ref{testwithalpha0,01and0,1and1}, the results of the maximum entropy method with different values of $\alpha$ are compared to the original spectral function. In Table 1, the 
error in the widths of the peaks in the spectral function are shown for the different values of $\alpha$, showing agreement within a few percent when $\alpha=1$.

\begin{table}
\begin{center}
	\begin{tabular}{|l|l|l|l|}
	\hline
	                  & $\alpha$=0.01 	&  $\alpha$=0.1 		&  $\alpha$=1		\\ \hline
	1st peak	& 17.15$\%$		& 14.84$\%$ 		& 3.58$\%$ 		\\ \hline
	2nd peak	& 22.85$\%$	 		& 10.29$\%$	 		& 4.65$\%$		\\ \hline
	\end{tabular}
\end{center}
\caption{The discrepency between the true spectral function and the results of deconvolution shown in Fig. \ref{testwithalpha0,01and0,1and1}.}
\end{table}

\begin{figure}
 \includegraphics[width=5in]{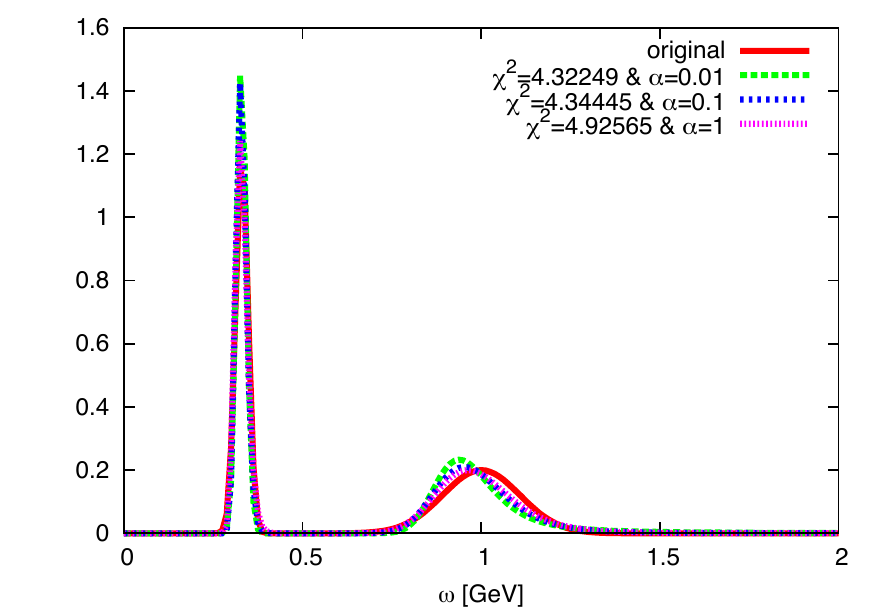}
\onlinecap
\caption{The spectral functions deconvolved from the test data  by using different values of $\alpha$. The corresponding values of $\chi^2$ are shown in the legend.
}
\label{testwithalpha0,01and0,1and1}
 \end{figure}

\begin{figure}
 \includegraphics[width=5in]{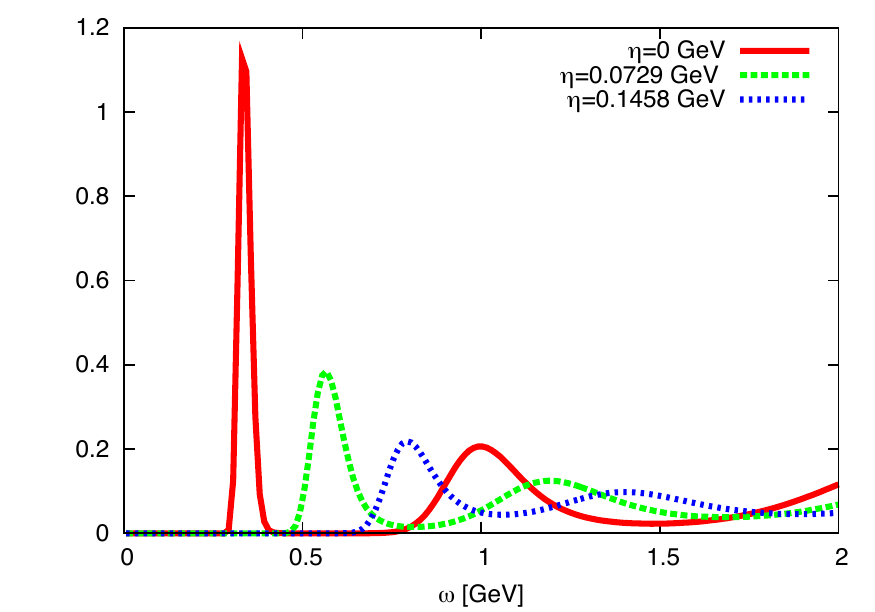}
\onlinecap
\caption{The spectral functions deconvolved from the results in Figure \ref{G_tau_tauShift2}. The corresponding values of $\chi^2$ are shown in the legend.
}
\label{rho_tauShift2}
 \end{figure}

\begin{figure}
 \includegraphics[width=5in]{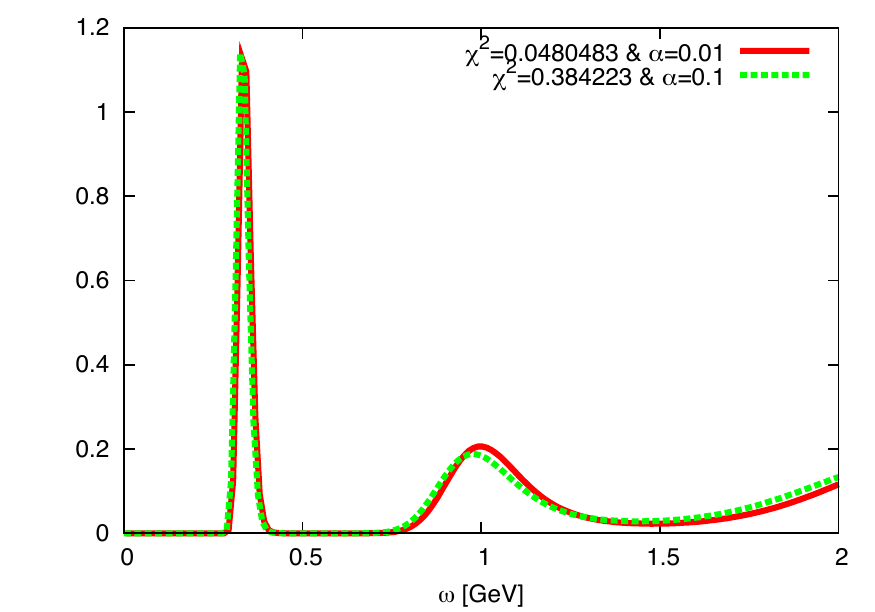}
\onlinecap
\caption{The spectral functions deconvolved from the results where $\eta=0$  in Figure \ref{G_tau_tauShift2} with different values of $\alpha$. The corresponding values of $\chi^2$ are shown
in the legend.
}
\label{quarkoniumspectralwithdifferentalpha}
 \end{figure}

The results from applying the maximum entropy method to the results for the quarkonium Green functions are shown in Figure \ref{rho_tauShift2}. 
It has resolved what should be a Dirac delta function in the spectral function without dissipation down to 
a width of 0.25 GeV. The spatial diffusion coefficient $2\pi TD = 5$ corresponds with an increase of the width to 1 GeV, signifying the state having a lifetime of 0.197 fm/c.
The results of the simulated annealing for 2$\pi$TD=$\infty$ with different $\alpha$ values can be seen in Figure \ref{quarkoniumspectralwithdifferentalpha}. It is observed that the results are not sensitive to the 
value of $\alpha$ over a considerable range. 

We end this section by noting the physics behind the changes of the spectral function in Figure \ref{rho_tauShift2}. First, the centroid of the ground state peak is shifted to high $\omega$ with increasing $\eta$. 
This suggests that the mean radius of the ground state increases, under the influence of momentum transfer from the medium. This could honestly have been seen rather quickly, by examining the slope of 
$G(\tau)$ in Figure \ref{G_tau_tauShift2}. A far less trivial result from the maximum entropy method is the increasing width of the ground state peak, indicating a decrease of the lifetime of the ground state. 

\section{Summary}

Deconvolution of spectral functions was examined with an emphasis on the building of intuition for quarkonium near $T_c$. Results from a treatment of quarkonium above deconfinement as an open quantum system yielded a set of spectral functions showing the effect of increasing $\eta$, and these results were found to be robust for a range of values in $\alpha$.

While this paper dug very deeply into the issues related to deconvolution, more work is required. In particular, only flat priors were used here; the role of the prior must be examined carefully. Finally, temperature-dependent potentials will be used in future work, making possible strong statements about quarkonium spectral functions including multiple temperature-dependent effects.

\section{Acknowledgments}

This work was supported by the Natural Sciences and Engineering Research Council of Canada and by the U.S. DOE Grant No. DE-FG02-87ER40328. We especially thank Kevin Dusling for many important comments at the earliest stages of this work.

\end{document}